\documentclass[aip,jcp,12pt,
preprint,
amsmath,amssymb,
floatfix,
]{revtex4-1}

\usepackage{graphicx}
\usepackage{dcolumn}
\usepackage{bm}

\begin{document}


\title{A resonance mechanism of efficient energy transfer mediated by Fenna-Matthews-Olson complex }

\author{Robert Alicki} 
\email{fizra@univ.gda.pl}
\affiliation{Institute of Theoretical Physics and Astrophysics, University
of Gda\'nsk,  Wita Stwosza 57, PL 80-952 Gda\'nsk, Poland}

\author{Wies{\l}aw Miklaszewski} 
\email{fizwm@univ.gda.pl}
\affiliation{Institute of Theoretical Physics and Astrophysics, University
of Gda\'nsk,  Wita Stwosza 57, PL 80-952 Gda\'nsk, Poland}

\begin{abstract}
The Wigner-Weisskopf-type model developed in [R. Alicki and F. Giraldi, J. Phys. B {\bf 44}, 154020 (2011)] is applied to the biological process of energy transfer from a large peripheral light harvesting antenna to the reaction center. This process is mediated by the Fenna-Matthews-Olson (FMO) photosynthetic complex with a remarkably high efficiency. The proposed model provides a simple resonance mechanism of this phenomenon employing exciton coherent motion and described by analytical formulas. A coupling to the vibrational environment is a necessary component of this mechanism as well as a fine tuning of the FMO complex Hamiltonian. The role of the relatively strong coupling to the energy sink in achieving the resonance condition and the absence of heating of the vibrational environment are emphasized.
\end{abstract}



\date{\today}
\maketitle
\section{Introduction}
The almost perfect energy transfer (APET) in photosynthesis phenomena, reaching the values over  95\%, attracts attention of physicists in the recent years \cite{May04}. In particular, the experimental observations of long-lasting quantum coherence in such systems \cite{Engel07,Scholes11} suggest that the popular picture of the excitation moving stepwise from the highest to the lowest excitonic state, by means of a diffusive hopping, need not to be correct. A natural consequence of this observation is that the quantum theory of open systems should be applicable to these phenomena. Indeed, a number of papers were published where the interplay between quantum coherence and environmental decoherence was considered as a plausible explanation of APET. Those papers present a variety of theoretical approaches which differ by a choice of mathematical modeling of an environmental noise. In some papers  Markovian models of Lindblad-Gorini-Kossakowski-Sudarshan are used \cite{Plenio08,Rebentrost09,Cao09, Hoyer10,Asadian10,Sarovar11,McCutcheon11,Jang11,Singh11,Ritschel11,Moix11,Scholak11,Schijven11}, in others  non-Markovian effects has been also included using hierarchical equations of motion \cite{Ishizaki09,Rebentrost11,Kreisbeck11} or certain "time-nonlocal" master equations \cite{Mohseni11}. In all those models reaction center plays a passive role of a probability sink. Different regimes can be observed, depending on the choice of parameters, and various optimization schemes are studied. The overall picture shows a noise-assisted enhancement of energy transport accompanied by suppression of  coherent oscillations, with the exception of \cite{Scholak11a}, where a small ensemble of molecular configurations yielding an efficient quantum transport (slightly reduced by Markovian dephasing), is found.
\par
 In the present paper a completely different  model is proposed -- a resonant, coherent energy exchange between the donor site and the acceptor one accompanied by an irreversible energy transfer to the sink (reaction center). The sink plays an active role modifying the energy of acceptor to achieve a resonance with respect to a donor. The applied mathematical formalism, based on the version of Wigner-Weisskopf model, developed in \cite{Alicki11} is used. In contrast to often very elaborate numerical analysis of master equations, or even the Schr\"odinger equation for the FMO complex and the bath, our approach employs simple analytical formulas which explicitly show a resonance APET mechanism and the necessity of fine-tuning of certain parameters. 
\section{A model of "bouncing exciton"}

The structure of monomeric subunit of the FMO complex (see \cite{Adolphs06}, Fig. 5) can be described  as a cavity made of proteins containing the BChl pigments and suggests the following analogy proposed in \cite{Alicki11}. Consider an optical cavity which consists of two parabolic mirrors with a common symmetry axis and two identical 2-level atoms placed in the focuses of the mirrors with transition dipoles parallel to the axis. A simple physical intuition suggests that when one of the atoms is excited the emitted photon will be bouncing between atoms during its lifetime which depends on the cavity quality factor. If additionally the first atom (donor) is excited by a light source (say a laser) and the second (acceptor) is coupled to an energy sink we obtain a mechanism of energy transfer from a source to a sink mediated by the donor-photon-acceptor  system.  The similar mechanism can be responsible for the efficient energy transfer in photosynthesis phenomena. Two outer BChl pigments coupled to the antenna and the reaction center, respectively, are analogs of two atoms -- donor and acceptor, and the other pigments together with the surrounding proteins form an environment which is the analog of electromagnetic field and supports wave-like traveling excitons -- the analogs of photons. Under certain resonance conditions, which for the optical analog are guaranteed by identical frequencies of both atoms, the symmetry of the cavity and the symmetry of atoms positions and orientations, the "bouncing exciton" can transport energy from the antenna to the reaction center. The efficiency of the process depends on two parameters $p$ and $q$. Here, $p$ is the probability that the energy which reaches the acceptor in "one-shot" is dissipated to the sink and $q$ is the probability of exciton recombination during the flight from the donor to the acceptor (or back). The simple analysis leads to the following formula for the efficiency of energy transfer after $n$ bounces 
\begin{equation}
\eta(n) = p(1-q) \frac{1-\bigl[(1-q)^2 (1-p)\bigr]^n}{1 - (1-q)^2 (1-p)}.
\label{bouncing}
\end{equation}
Using, the relation $q\simeq t_0/\tau$ where $t_0 \simeq 1$ps is a typical time of flight and $\tau\simeq 1$ns is a typical exciton recombination time one obtains $q\simeq 10^{-3}$. In the next Sections we show that under reasonable conditions $p\simeq 0.5$. Hence, using the above choice of parameters we obtain the following estimations for the "five-steps" and asymptotic efficiencies
\begin{equation}
\eta(5)\simeq 0.97,~~~\eta(\infty) \simeq 1-q\bigl[1+ \frac{2(1-p)}{p}\bigr] \simeq 0.997
\label{bouncing1}
\end{equation}
which are sufficiently high to account for the observed in Nature values over 0.95.

\section{ A model of excitonic energy transport}

We consider a standard  tight-binding model of  energy transport in quantum networks in the single-exciton approximation which is valid under the assumption that the exciton lifetime is much longer than any other relevant time scale in this model. The initial "skeleton system" consists of $N+1$ sites with the corresponding single-exciton Hilbert space ${\mathbb C}^{N+1}$ and the Hamiltonian matrix $H = [h_{kl}; k,l = 1,2,...,N,N+1]$. The following assumptions,  notation 
and terminology will be used:
\begin{itemize}
\item The diagonal elements $h_{kk}$ are energies of the sites (denoted by $\omega_k = h_{kk}$) while $\{h_{kl}; k \neq l\}$ are hopping amplitudes.
\item The sites "1" , "2" , "N+1" correspond to a donor, an acceptor and a sink, respectively.
\item The direct coupling between the donor and the acceptor is usually small because of their spatial separation and similarly the sink is strongly coupled only to the acceptor, so we can put
\begin{equation}
h_{12} = 0,~~~h_{j,N+1} = 0~~~~\mathrm{for}~~~ j\ne 2
\label{coupling}
\end{equation}
what essentially simplifies the model.
\end{itemize}
Under these conditions the  Hamiltonian can be recasted into the form which reflects the open system character of the model
\begin{equation}
H = \omega_1 |1\rangle\langle 1| + \omega_2 |2\rangle\langle 2| + H_B + \left(|1\rangle\langle g_1| + |2\rangle\langle g_2|+\mathrm{h.c.}\right) .
\label{discreteWW}
\end{equation}
Here a donor and an acceptor form a system with the Hamiltonian given by the first two terms of (\ref{discreteWW}), the "skeleton bath" consists of the rest $N-1$ sites including a sink with the Hamiltonian $H_B$. The last  terms describe the system-bath coupling,  
$|1\rangle =[1,0,...,0]$, $|2\rangle =[0,1,...,0]$, $|g_1\rangle = [0,0,h_{13},...h_{1N},0]$, $|g_2\rangle = [0,0,h_{23},...h_{2N}, h_{2,N+1}]$ and $H_B$ is a submatrix of $H$ with indices  $k , l = 3, 4,...,N, N+1$, namely 

\begin{eqnarray}
H&=\left(\begin{array}{cccccc}
\omega_1 & 0 &h_{13} & \cdots & h_{1N}&0\\
0 &\omega_2 & h_{23} & \cdots & h_{2N}&h_{2,N+1}\\
h_{13}^\ast &  h_{23}^\ast &\omega_3 &  \cdots & h_{2N}&0\\
\vdots &  \vdots &\vdots &  \ddots & \vdots &0\\
h_{1N}^\ast&  h_{2N}^\ast &h_{3N}^\ast &  \cdots &\omega_N &0\\
0&  h_{2,N+1}^\ast &0 &  \cdots&0 &\omega_{N+1}
\end{array}
\right).
\label{ham}
\end{eqnarray}
$H_B$ can be written in its spectral decomposition form
\begin{equation}
H_B = \sum_{\alpha =3}^{N+1}\epsilon_{\alpha} |\alpha\rangle\langle\alpha | 
\label{specH_1}
\end{equation}
where $|N+1\rangle = [0,...0,1]$ and $\epsilon_{N+1}\equiv \omega_{N+1}$ correspond to the sink whose energy should be the lowest one and well-separated from the others. 

However, this model with the low dimensional Hilbert space and hence the Hamiltonian with well-separated energy levels cannot describe irreversible energy transfer phenomena. We have to take into account the coupling to vibrational degrees of freedom. This interaction transforms the eigenstates  $|\alpha\rangle$  into resonances with finite spectral widths (decoherence rates) $\Gamma_{\alpha}$. The finite-dimensional Hilbert space spanned by $\{|\alpha\rangle\}$ is replaced by an infinite-dimensional $L^2({\mathbb R})$ and the Hamiltonian (\ref{specH_1}) by the continuous spectrum multiplication operator by $\omega$. We do not need to determine the exact form of the total Hamiltonian denoted by $H'$
but the whole information will be included in the form of the relevant correlation functions \cite{Alicki11}.
The system begins its evolution in the donor state $|1\rangle $ and after time $t$ can be found in the acceptor state $|2\rangle$ with the probability
\begin{equation}
\mathcal{P}_{12}(t) = |\mathcal{A}_{12}(t) |^2,~~~ \mathcal{A}_{12} =\langle 2|e^{-iH't} |1\rangle.
\label{Pro}
\end{equation}
The APET takes place if  for a certain time $t_0$ the transfer probability $\mathcal{P}_{12}(t_0)\simeq 1$. Then the energy must be transferred from the acceptor to the sink in an irreversible process
which is fast in comparison to the lifetime of the acceptor state.
\par
The relevant correlation functions $\langle g_j|g_j(t)\rangle$  are now given by 
\begin{equation}
G_{jj}(t) = \sum_{\alpha =3}^{N+1} |\langle\alpha|g_j\rangle|^2 e^{-i\epsilon_{\alpha}t}e^{-\Gamma_{\alpha}t},~~~~j=1,2, ~~~ t\geq 0 \label{corr1}
\end{equation}
where the presence of decoherence rates $\Gamma_{\alpha}$ reflects the fact that the states $|\alpha\rangle$ are not the eigenstates of the modified Hamiltonian $H_B'$ but the unstable resonances embedded into practically continuous spectrum.
\par
The Laplace transforms of the correlation functions read
\begin{equation}
 \int_0^{\infty} e^{i\omega t}G_{jj}(t)\,dt = \gamma_j(\omega) + i\,\delta_j(\omega)
\label{laplace}
\end{equation}
with the decay rates
\begin{equation}
\gamma_j(\omega) =  \sum_{\alpha =3}^{N+1} |\langle\alpha|g_j\rangle|^2\frac{\Gamma_{\alpha}}{(\omega-\epsilon_{\alpha})^2 + \Gamma_{\alpha}^2}
\label{decay}
\end{equation}
and the energy shifts
\begin{equation}
\delta_j(\omega) =  \sum_{\alpha =3}^{N+1} |\langle\alpha|g_j\rangle|^2\frac{\omega -\epsilon_{\alpha}}{(\omega-\epsilon_{\alpha})^2 + \Gamma_{\alpha}^2}  .
\label{shift}
\end{equation}
The final approximate formula for the optimal transfer probability based on the derivation presented in  \cite{Alicki11} which is valid under the condition that the bath-mediated interaction  and the collective dissipation effects  between the donor and the acceptor can be neglected, reads
\begin{equation}
\mathcal{P}_{12}(t_0) \simeq \left|\int e^{i\theta(\omega ; t_0)} \sqrt{f_1(\omega) f_2(\omega)}d\omega\, \right|^2.
\label{approx1}
\end{equation}
Here
\begin{equation}
f_j(\omega) = \frac{1}{\pi}\frac{\gamma_j(\omega)}{\left[\omega - \omega_j - \delta_j(\omega)\right]^2 +\gamma_j(\omega)^2},~~~~ j=1,2
\label{approx2}
\end{equation}
are interpreted as normalized \cite{Norm} moduli squares of the wave packets emitted to the bath from the donor (forwards in time) and from the acceptor (backwards in time), respectively.
The oscillating term  $\exp[i\theta(\omega ; t_0)]$ containing the relative phase of these two wave packets reduces the value of transition probability and will be discussed in the next Section. 

\section{ The role of  sink and  phase-factor }
In order to give a more transparent picture of APET we approximate the functions $f_j(\omega)$ in (\ref{approx2}) by  Lorentzians 
\begin{equation}
f_j(\omega) = \frac{1}{\pi}\frac{\gamma_j}{\left(\omega - \omega^r_j \right)^2 +\gamma_j^2}
\label{lorentz}
\end{equation}
where $\gamma_j = {\gamma_j}(\omega^r_j)$
and the renormalized frequencies satisfy the following self-consistent equation
\begin{equation}
\omega^r_j = \omega_j + \delta_j(\omega^r_j).
\label{lorentz2}
\end{equation}
The APET condition means the perfect overlap of both Lorentzians what implies
\begin{equation}
\omega^r_1 = \omega^r_2 = \omega_0,~~~~\gamma_1 = \gamma_2 =\gamma.
\label{apet}
\end{equation}
Notice that due to the condition $h_{1,N+1}\equiv \langle N+1|g_1\rangle =0$ the renormalized frequency $\omega^r_1$ and the decay rate $\gamma_1$ of the donor do not depend on the sink parameters $h_{2,N+1}\equiv \langle N+1|g_2\rangle $, $\epsilon_{N+1}$ and $\Gamma_{N+1}$. Therefore, the APET conditions (\ref{apet}) can be realized by fitting the sink parameters to modify properly the renormalized frequency $\omega^r_2$ and the decay rate $\gamma_2$ of the acceptor. The resonance condition (\ref{apet}) implies that the exciton energy at the donor is the same as at the acceptor and equal to $\hbar\omega_0$ what means the absence of heating of the vibrational environment. To achieve a rapid and irreversible energy transfer from the acceptor to the sink, the acceptor-sink coupling $h_{2,N+1}$ and the decoherence rate for the sink $\Gamma_{N+1}$ should be much larger than the decay rate $\gamma$.
Moreover, the parameter $\Gamma_{N+1}$ can be treated as a sum of two terms,  $\Gamma_{N+1}=\Gamma_{N+1}^d +\Gamma_{N+1}^r$ where $\Gamma_{N+1}^d$ corresponds to the decoherence phenomena while $\Gamma_{N+1}^r$ describes exciton decay in the sink. In fact one should add $-i \Gamma_{N+1}^r$ to the sink energy $\epsilon_{N+1}$ and then transform the initial Hamiltonian into a nonhermitian one to include this decay process. Such interpretation does not change the basic formula (\ref{approx1}) at resonance as the imaginary part of the phase $\theta(\omega ; t_0)$, due to $-i \Gamma_{N+1}^r$, appears around the sink energy $\epsilon_{N+1}$ which is assumed to be far away from the resonance value $\omega_0$. On the other hand it provides an explanation of the energy transfer irreversibility.
\par
Finally, one has to discuss the influence of the oscillating factor which under the conditions (\ref{lorentz}) and (\ref{apet}) possesses a much simpler structure derived using the results of \cite{Alicki11}
\begin{equation}
\theta(\omega ; t_0) = 2\arctan \left(\frac{\omega - \omega_0}{\gamma}\right)- \left(\tau(\omega) - t_0\right)(\omega - \omega_0), 
\label{phase}
\end{equation}
where we use the parametrization of the phase-factor in terms of $\tau(\omega)$ which can be interpreted as the propagation time of the exciton wave packet, with frequencies concentrated around $\omega$,  between the donor and the acceptor. For a constant exciton velocity, i.e., $\tau(\omega) = \tau$ an optimal choice of $t_0 = \tau + 1/\gamma$
gives the transition probability $4/e^2\approx 0.54$. Simple numerical estimations show that for the choice $\tau(\omega)= \tau(\omega_0)
+ \kappa (\omega - \omega_0)^2 $ one can reach probability $0.72$. These relatively low probabilities are caused by the process of exciton back-scattering by the acceptor coupled to the sink. In order to explain the observed efficiencies higher than  $0.95$ we need the "bouncing exciton" model described above. In the following we shall concentrate on the values of the overlap integral
\begin{equation}
\mathcal{F}=\left[\int \sqrt{f_1(\omega) f_2(\omega)}\, d\omega\right]^2 .
\label{efficiency}
\end{equation}
For example, if $\mathcal{F}\simeq 0.7$ and the presence of the phase (\ref{phase}) introduces another factor $0.7$, one can estimate the single-shot transfer probability  as $p\simeq 0.5$ which is high enough to account for the observed phenomena.
\section{ Results for FMO complex}
We apply the presented theory to the case of the FMO complex of \emph{C. tepidum} using the data of \cite{Hoyer10,Adolphs06}. The model of FMO complex is given by the Hamiltonian of type (\ref{ham}) with $N=7$ containing $7\times 7 $ Hamiltonian submatrix taken from \cite{Adolphs06}. We adopt a standard choice of the BChl 1 and BChl 3 molecules as the donor and acceptor, respectively. Notice, that in the recent literature \cite{Busch11,Ritschel11,Olbrich11,Moix11} a new model containing eight BChl pigments is applied, however, with a still large variance between Hamiltonian parameters found in the literature. As the aim of the present work is not to give the ultimate quantitative description, but rather to illustrate the new mechanism of energy transfer, we use the "old model".

The free parameters of the model  are randomly sampled within the following intervals: $\omega_8 \in [-500,0]$, $\Gamma_j \in [50,90]$, $j= 1,\ldots,8$ (all values in cm$^{-1}$). The limits put on the decay rates $\Gamma_j$, $j= 1,\ldots,8$, are consistent with the experimental widths of absorption spectral lines obtained at the temperature $T_0= 77 K $ \cite{Cho05}. The bound for $\omega_8$ is comparable to the largest energy variation in the FMO complex and the value of $|h_{28}|$ is practically unrestricted. Notice, that the value of $\omega_8$ must be negative, in order to "push up" acceptor energy to achieve resonance with a donor.

\begin{figure}[!ht]
\begin{center}
\includegraphics[width=0.45\textwidth]{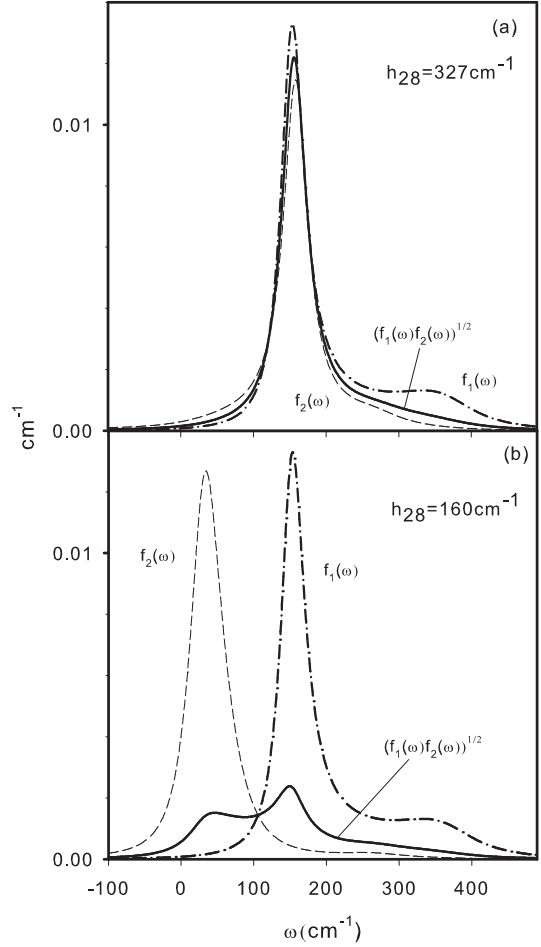}
\end{center}
\caption{The functions $f_i(\omega)$ and  $\sqrt{f_1(\omega) f_2(\omega)}$  for optimized parameters (in cm$^{-1}$): $\Gamma_3$= 59.6, $\Gamma_4$= 90.0, $\Gamma_5$= 50.3, $\Gamma_6$= 59.7, $\Gamma_7$=89.7, $\Gamma_8$=50.1, $h_{28}$=327 and $\omega_8$=-500 giving $\mathcal{F}\approx 0.75$ (a) and with shifted $h_{28}=160$  resulting in $\mathcal{F}\approx 0.18$ (b).}
\label{fig1}
\end{figure}

\begin{figure}[!ht]
\begin{center}
\includegraphics[width=0.45\textwidth]{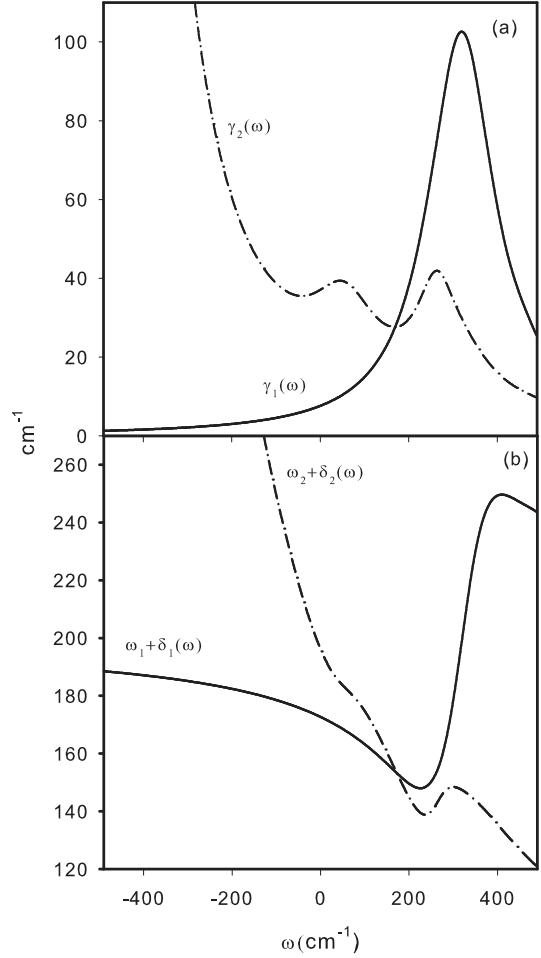}
\end{center}
\caption{The functions   $\gamma_i(\omega)$ (a) and $\delta_i(\omega)$ (b) for optimized parameters (see Fig. \ref{fig1}) giving $\mathcal{F}\approx 0.75$.}
\label{fig2}
\end{figure}

\begin{figure}[!ht]
\begin{center}
\includegraphics[width=0.5\textwidth]{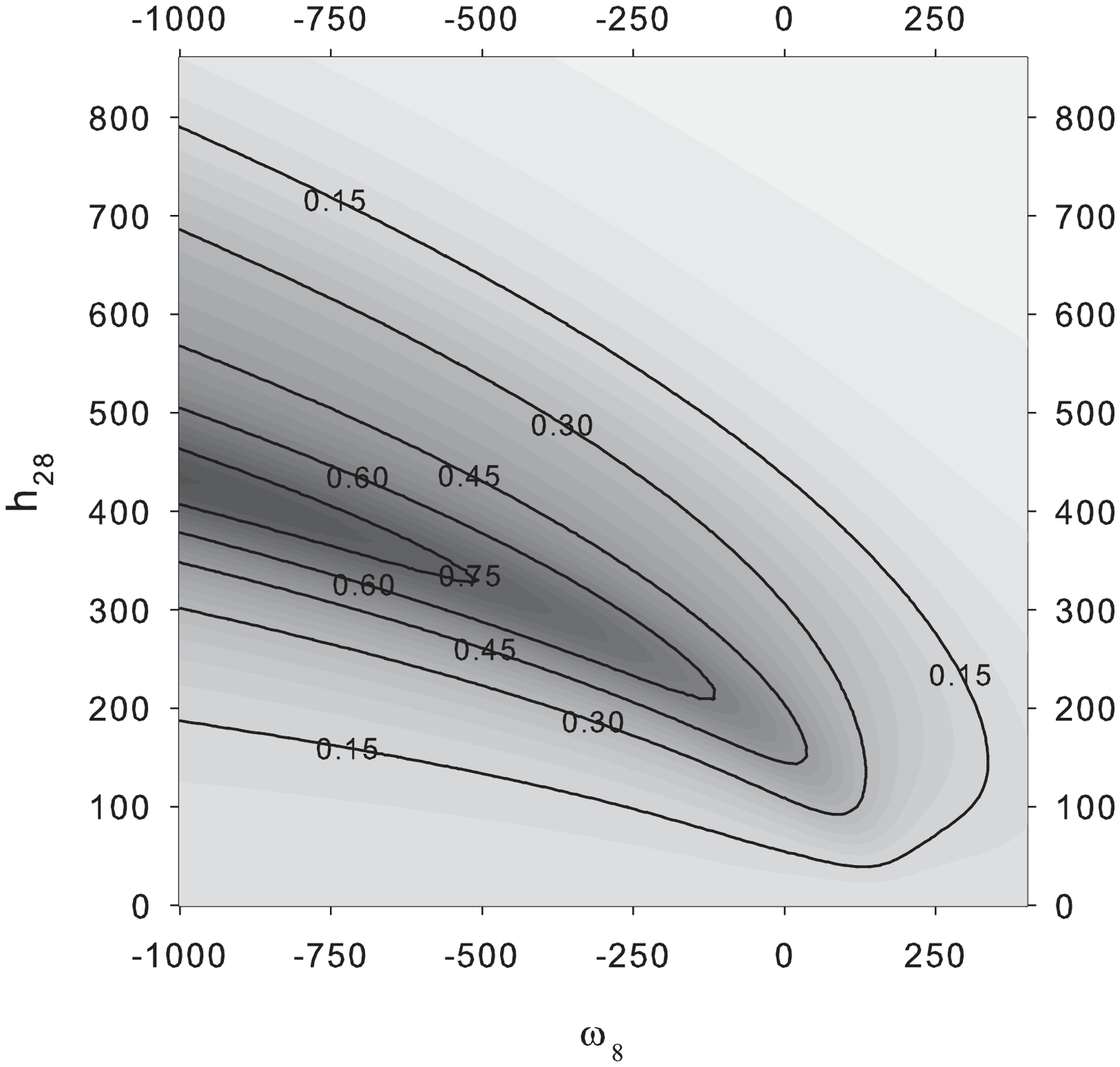}
\end{center}
\caption{ The factor $\mathcal{F}$ vs. $|h_{28}|$ and $\omega_8$ for optimized rates $\Gamma_i$ (see Fig. \ref{fig1}).}
\label{fig3}
\end{figure}

Under these conditions the maximal value of the factor
$\mathcal{F}\simeq 0.75$ and the corresponding resonance phenomenon at the point $\omega_0 \simeq 150$cm$^{-1}$ is clearly visible (see Fig.  \ref{fig1}a) with the width $\gamma\simeq 30$cm$^{-1}$. The interesting feature is that the curves representing the shifted energies for the donor and the acceptor are very close in a rather wide interval of frequencies
 $[100, 300]$ (Fig. \ref{fig2}b). A similar behavior is observed for decay rates as well (Fig. \ref{fig2}a). It suggests that the Hamiltonian of FMO complex is well-tuned to make the resonance condition less sensitive with respect to the parameters $|h_{28}|$ and $\omega_8$. The efficiency dependence on these parameters is illustrated by Fig. \ref{fig3} and in Fig. \ref{fig1}b. 

The predictions concerning the sink parameters are the following: 
\begin{enumerate}
\item The ratio $|\omega_{8}|/ |h_{28}| \simeq 1.5$.
\item The value of $|h_{28}|$ or $-\omega_8$ should reach  maximum available by the physical mechanisms in such systems.
\end{enumerate}
For our choice of the limit for $\omega_8$ the optimal value of acceptor-sink coupling $|h_{28}|$ is about three times larger than the largest value of the dipole-dipole coupling between pigments in the FMO complex. This can be  explained  by the expected close contact between the acceptor and the reaction center \cite{Wen09}.

\begin{figure}[!ht]
\begin{center}
\includegraphics[width=0.5\textwidth]{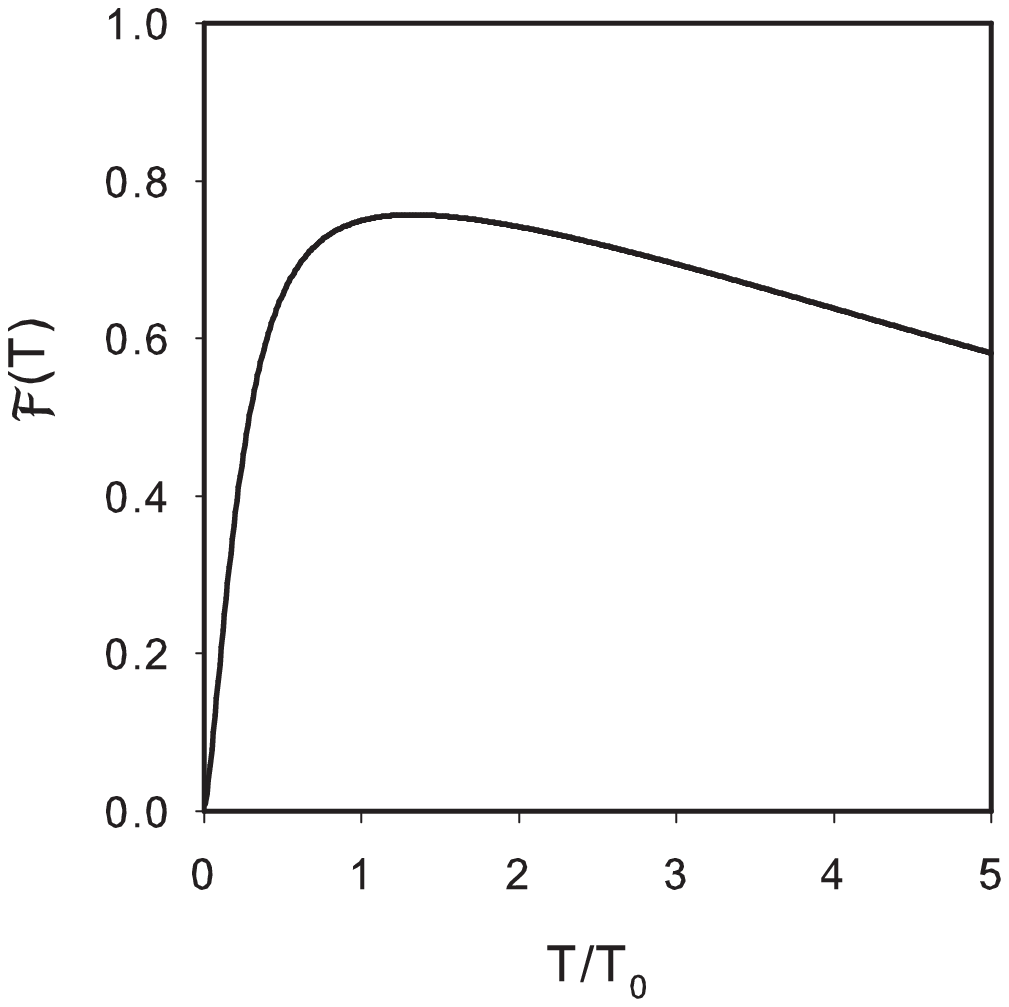}
\end{center}
\caption{ The temperature dependence of the factor $\mathcal{F}$ for optimized parameters (see Fig. \ref{fig1}).}
\label{fig4}
\end{figure}

To illustrate the temperature dependence of $\mathcal{F}(T)$ we assume that all decay rates are proportional to the environmental temperature \cite{Temp} and hence $\Gamma_k(T) = \Gamma_k(T_0)\times (T/T_0) , k=1,\ldots,8$. The results, with the optimal choice of $\Gamma_k(T_0)$ as in Fig. \ref{fig1}, are illustrated in Fig. \ref{fig4}.

\section{ Concluding remarks} 

In contrast to the models based on Markovian or non-Markovian relaxation towards the lower energy levels in our model the exciton energy is not dissipated into vibrational environment of the FMO complex but is transfered in a resonant process to the acceptor and then transformed into other forms of energy in the reaction center. This implies the absence of heating of the vibrational degrees of freedom, which is considerably large in the standard models. Namely, in those models the difference between donor and acceptor energy is dissipated into environment before the exciton can reach the reaction center. This portion of energy  $\sim 200$cm$^{-1}$ $\sim k_B\times 300K$ per exciton, can have biological relevance and, perhaps, could be used as a test for the proposed models. On the other hand, the vibrational degrees of freedom provide the necessary "quasi-continuous medium" for the energy propagation. The process of energy transfer is a coherent one and its efficiency is increased by the mechanism of "bouncing". Temperature dependence of the efficiency shows its rapid enhancement with increasing temperature and the relative stability in the vast region of high temperatures. The fine-tuning concerns  mainly the Hamiltonian of the FMO complex
what suggests an evolutionary natural selection mechanism. All these features seem to be consistent with the phenomenology of energy transfer in the FMO complex.

\acknowledgments

R. A. acknowledges the Polish Ministry of Science and Higher Education, grant \\ PB/2082/B/H03/2010/38.


\begin{thebibliography}{29}
\bibitem [{Nor()}]{Norm}%
  \BibitemOpen
  \href@noop {} {}\bibinfo {note} {Numerical calculations suggest that the
  function (\ref{approx2}) is always normalized to one, but the mathematical
  proof is missing.}\BibitemShut {Stop}%
\bibitem [{Tem()}]{Temp}%
  \BibitemOpen
  \href@noop {} {}\bibinfo {note} {This can be derived assuming that the
  eigenstates $|\alpha\rangle$ are replaced by ground states of certain
  fictitious oscillators coupled to heat baths at the temperature
  $T$.}\BibitemShut {Stop}%
\end{thebibliography}

%

\end{document}